\documentclass{article}
\usepackage{amsfonts}
\usepackage{spconf,amsmath,graphicx,epsfig}
\usepackage{graphicx}
\usepackage{color}
\usepackage{mathtools}
\usepackage{microtype}
\usepackage{xcolor}
\usepackage{caption}
\usepackage{subcaption}
\usepackage{lineno}
\usepackage{tabularx}
\usepackage{units}
\usepackage{multirow}

\usepackage{array, tabularx,  ragged2e,  booktabs}

\title{Scaling Up Music Information Retrieval Training with \\  Semi-Supervised Learning}

\name{Yun-Ning Hung$^{1}$, Ju-Chiang Wang$^{1}$, Minz Won$^{1}$, Duc Le$^{1}$}
\address{$^1$ SAMI, ByteDance, San Jose, CA, USA}

\def\authorname{Y.-N. Hung, J.-C. Wang, M. Won, D. Le}

\begin{document}

\maketitle
\begin{abstract}
In the era of data-driven Music Information Retrieval (MIR), the scarcity of labeled data has been one of the major concerns to the success of an MIR task. In this work, we leverage the semi-supervised teacher-student training approach to improve MIR tasks. For training, we scale up the unlabeled music data to 240k hours,
%pull music data from a variety of sources and scale up the training data to 240k hours, 
which is much larger than any public MIR datasets. We iteratively create and refine the pseudo-labels in the noisy teacher-student training process. Knowledge expansion is also explored to iteratively scale up the model sizes from as small as less than 3M to almost 100M parameters. We study the performance correlation between data size and model size in the experiments. By scaling up both model size and training data, our models achieve state-of-the-art results on several MIR tasks compared to models that are either trained in a supervised manner or based on a self-supervised pretrained model. To our knowledge, this is the first attempt to study the effects of scaling up both model and training data for a variety of MIR tasks.
\end{abstract}


%\vspace{-0.4\baselineskip}

\section{Introduction}\label{sec:introduction}
Large-scale datasets together with increasing model sizes have been recognized as one of the critical factors to build strong machine learning systems. This direction has led to rapid success in many domains, such as natural language processing~\cite{OpenAI2023GPT4TR},
%, Smith2022UsingDA}
computer vision~\cite{Goyal2021SelfsupervisedPO},
%, Zhai2021ScalingVT}
and automatic speech recognition (ASR)~\cite{Xiao2021ScalingAI}. 
%Zhang2021BigSSLET,
Plenty of resources from Internets and publicly available datasets enable the models to scale up to billions of parameters and achieve state-of-the-art results on several downstream tasks. 

In the domain of music information retrieval (MIR), the lack of labeled datasets has hindered the advances of several downstream tasks \cite{Won2021SemisupervisedMT, Simon2022ScalingPT}. Although large-scale datasets such as MSD~\cite{Bertin-Mahieux2011} and Jamendo~\cite{Bogdanov2019TheMD} have contributed to significant progress in music auto-tagging, they are still considered smaller-scale datasets in the aforementioned domains. Other MIR tasks, such as chord recognition and structure segmentation, have an even smaller amount of training data due to the labeling difficulty. For example, labeling time-varying musical events at a millisecond scale such as beats requires specialized training and is very labor-intensive.

To tackle this problem, several prior works~\cite{Dhariwal2020JukeboxAG, Li2023MERTAM} have investigated leveraging self-supervised approaches to train on large-scale data and transfer the knowledge to downstream tasks. Castellon et al.~\cite{Castellon2021CodifiedAL} has shown that probing the pretrained Jukebox model can achieve better results in emotion recognition and comparable results in music auto-tagging compared to supervised approaches. Li et al.~\cite{Li2023MERTAM} demonstrated that using large-scale data to train a large-scale self-supervised model, the pretrained MERT can outperform the supervised setting in several downstream tasks. 
%Despite their promising results, 
Due to the characteristics of different MIR downstream tasks, pretrained models have not been observed to outperform some supervised approaches for tasks such as key detection. %We will also show in Section~\ref{sec:result} that both Jukebox and MERT do not perform well for tasks that required detailed harmonic information, such as chord recognition and downbeat tracking. 

\begin{table}
\small
\noindent\setlength\tabcolsep{4pt}%
\begin{tabularx}{\linewidth}{>{\RaggedRight\arraybackslash\hsize=.80\hsize}X >{\RaggedRight\arraybackslash\hsize=.20\hsize}X}
  \toprule
  Data Sources & Hours    \\ 
  \midrule 
  RWC \cite{goto2002rwc}, Ballroom \cite{gouyon2005computational}, Beatles \cite{davies2009evaluation}, USPOP \cite{Berenzweig2004ALE}, Billboard \cite{smith2011design}, Hainsworth \cite{gouyon2006experimental}, SMC \cite{gouyon2005computational},
  Isophonics \cite{mauch2009omras2}, HookTheory \cite{donahue2022melody},
  HJDB \cite{hockman2012one},
  SALAMI \cite{smith2011design}, Simac \cite{holzapfel2012selective}, HarmonixSet\cite{nieto2019harmonix},  Internal-Structure, %\cite{wang2021supervised}, 
  Internal-Beat/Key %\cite{heydari2023singnet}
  & ~1000\\
  \midrule 
  Audio-92K & 92,000\\
  \midrule 
  Music-151K & 151,000\\
  \bottomrule
\end{tabularx}
\captionof{table}{Statistics of each data source.} 
 \label{tab:supervised_datasets}
\vspace{-0.2cm}
\end{table}

In this work, we attempt to scale up the training data and model sizes from a different angle. We leverage the semi-supervised learning to train on both labeled and unlabeled datasets. To scale up the model size, we use the training framework of teacher-student knowledge expansion to iteratively expand the size of student models. Teacher-student training has demonstrated its usefulness in several MIR tasks ~\cite{Won2021SemisupervisedMT, Hung2022ALT, Simon2022ScalingPT}. We push this idea to a new limit by expanding the models from as small as less than 3 million to almost 100 million parameters. To scale up the data size, we collect labeled datasets for various MIR tasks and mix them together, as shown in Table~\ref{tab:supervised_datasets}. To further expand the training data, two internal datasets (with 1.3 million and 2.4 million full songs respectively) are included, yielding 300 times more than any other publicly available datasets (e.g. Jamendo \cite{Bogdanov2019TheMD} only has 3760 hours) in the MIR domain.

In the experiments, we investigate the performance correlation between data size and model size. We test the proposed method on four MIR tasks and show that by scaling up both model and training data, the proposed approaches can outperform not only existing self-supervised approaches but also achieve state-of-the-art results on these MIR tasks. 

% \vspace{0.8\baselineskip}

\begin{table*}
\centering
\small
\begin{tabular}{c c c c c c c c c c c c}
  \hline
   & Round & Params & $R$ & Pool & Hop & Layer & Spec D. & Temp D. & Heads & L. Arr & Augment \\
  \hline \hline
  \multirow{3}{*}{Beat} & 1st  & 11M & \multirow{3}{*}{1} & \multirow{3}{*}{(2, 2)} & \multirow{3}{*}{160}   & 5 & 32  & 128 & 4 & 4 & A1, A2\\\
                           & 2nd & 43.3M & & &  & 5 & 64  & 256  & 4 & 4 & A1, A2, A4\\
                           & 3rd & 91.3M  & & &  & 6 & 256 & 256  & 4 & 4 & A1, A2, A4\\
  \hline
  \multirow{3}{*}{Chord} & 1st  & 2.9M & \multirow{3}{*}{2} & \multirow{3}{*}{(4, 4)} & \multirow{3}{*}{500}  & 5 & 64  & 64 & 4 & 8 & A1, A2\\\
                           & 2nd & 11.4M & & &  & 5 & 128  & 128  & 4 & 8 & A1, A2, A3\\
                           & 3rd & 91.9M  & & &  & 6 & 256 & 256  & 8 & 8 & A1, A2, A3\\
  \hline                    
  \multirow{3}{*}{Key} & 1st  & 1.5M & \multirow{3}{*}{3} & \multirow{3}{*}{(4, 100)} & \multirow{3}{*}{320}  & 5 & 128  & 32 & 4 & 4 & A1, A2\\
                           & 2nd & 5.9M & & &  & 5 & 256  & 64  & 4 & 4 & A1, A2, A3\\
                           & 3rd & 42.8M  & & &  & 8 & 512 & 128  & 8 & 8 & A1, A2, A3\\
  \hline
  \multirow{3}{*}{Structure} & 1st  & 2.9M & \multirow{3}{*}{1} & \multirow{3}{*}{(4, 6)} & \multirow{3}{*}{512}  & 5 & 64  & 64 & 4 & 8 & A1, A2\\\
                           & 2nd & 11.4M & & &  & 5 & 128  & 128  & 4 & 8 & A1, A2, A3\\
                           & 3rd & 91.9M  & & &  & 6 & 256 & 256  & 8 & 8 & A1, A2, A3\\
  \hline
 \end{tabular}
 \caption{Parameters of each task for three rounds of training. ``Pool'': the maxpooling sizes (F, T) for frequency and time axes. ``Hop'': hop size. ``Spec D.'' and ``Temp D.'': the latent dimensions for spectral and temporal Transformers, respectively \cite{Lu2023MultitrackMT}. ``Heads'': number of  attention heads. ``L. Arr'': number of latent arrays.  ``Augment'': augmentation methods (see Section~\ref{sec:aug}).} 
 \label{tab:parameters}
\end{table*}

%\vspace{-0.8\baselineskip}

\section{Methodology}

\subsection{Training Scheme}
We use the noisy student training following previous works~\cite{Xie2019SelfTrainingWN}. The learning process includes four steps:
\vspace{-0.1cm}
\begin{enumerate}
  \itemsep-0.2em 
  \item Train a supervised teacher $M$ using small training set. 
  \item Use the trained teacher model to assign pseudo-labels to unlabeled samples. 
  \item Apply data augmentation to the audio of pseudo-labeled samples, then mix them with the supervised samples to create a new training set.
  \item Train a student model with the same or larger size using the new training set.
\end{enumerate}
\vspace{-0.2cm}
After the student model is well-trained and outperforms the teacher, we replace the old teacher with the student and repeat steps 2 -- 4 to train a new student model.  

\begin{figure}
\centering
\includegraphics[width=0.7\linewidth]{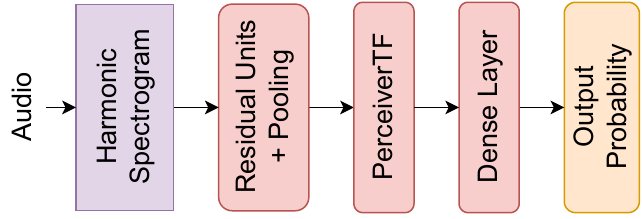}
\caption{Model architecture overview.}
\label{fig: framework}
\end{figure}

%\vspace{-0.6\baselineskip}

\subsection{Model architecture}

Fig.~\ref{fig: framework} depicts the proposed model architecture. The input audio is first transformed into spectrogram.
%Following previous works~\cite{Won2020DataDrivenHF, Hung2022ModelingBA}, 
We use six trainable harmonic filters~\cite{Won2020DataDrivenHF, Hung2022ModelingBA} to enhance the spectrogram. Then, $R$ residual units~\cite{he2016identity} followed by a stack of spectro-temporal max-pooling layers are used to produce a time-frequency representation.
% downsample the frequency/temporal resolution.  ### it increase channel size significantly, so in total it's not downsampled
Each residual unit has a kernel size of 3 and $F$ feature maps, where $F$ is the spectral dimension in Table~\ref{tab:parameters}. 
We use PerceiverTF~\cite{Lu2023MultitrackMT}, one from the Transformer family, as the %main body of our 
model architecture. PerceiverTF has shown state-of-the-art performance in multi-instrument music transcription~\cite{Lu2023MultitrackMT}, demonstrating its ability to characterize useful information for modeling both pitch- and time-related events. More importantly, PerceiverTF incorporates the Perceiver structure \cite{jaegle2021perceiver} to improve training scalability, making it a good fit for this work over SpecTNT~\cite{Lu2021SpecTNTAT}. We use exactly the same Perceiver TF blocks as in ~\cite{Lu2023MultitrackMT}. 
% Due to the success of Transformer architecture in several domains, a transformer, Perceiver~\cite{Lu2023MultitrackMT}, is concatenated after to extract high-level embedding. 
% We use Perceiver since it has the ability to capture both time and frequency information but at the same time improve the training efficiency compared to SpecTNT~\cite{Lu2021SpecTNTAT}. 
Lastly, a dense layer is added to predict the frame-level labels from the PerceiverTF output.
In each training round, we gradually scale up the parameters of PerceiverTF until it is close to 100M, as shown in Table~\ref{tab:parameters}. 

\subsection{Data Augmentation} \label{sec:aug}
We use the following augmentations to train the noisy student:  
%are considered to increase the difficulty of student training:
\begin{enumerate}
\vspace{-0.2cm}
  \itemsep-0.2em 
  %\item  Time stretching: as described in ~\cite{Bck2020DeconstructAR}, we randomly change up to $\pm 20 \%$ of the hop size to modify the tempo.
  %\item Pitch shifting: since we use constant-Q transform to get the spectrogram, each frequency bin corresponds to one semitone. Hence, we randomly shift the pitches by rotating the frequency bins in a range of $\pm 3$ semitones in the experiments.
  \item Pitch shifting (A1): we randomly shift the pitches in a range of $\pm 3$ semitones in the experiments.
  \item Torchaudio\_augmentations\footnote{https://github.com/Spijkervet/torchaudio-augmentations} (A2): with a probability $p=0.8$, we add audio transformations including white noise (SNR in the range between 0.3 and 0.5 dB), random gains, high/low pass filters, and polarity inversion.
  \item Frequency masking (A3): we apply two frequency channels masking blocks; each randomly masks out 0--25\% of the input frequency bins.
  \item Time masking (A4): we apply two time steps masking blocks; each randomly masks out 0--25\% of the frames.
\end{enumerate}
\vspace{-0.1cm}
\noindent We use torchaudio library\footnote{https://pytorch.org/audio/stable/transforms.html} for the frequency and time masks.

\section{Data Preparation}

\subsection{Data Sources}

As summarized in Table~\ref{tab:supervised_datasets}, we used the datasets in the first row for supervised training. Except for publicly available datasets, we include two in-house datasets for supervised training for two reasons. First, we want to maximize the size of labeled datasets, as Transformer-based architecture is data-hungry. Second, a strong teacher model can build a better foundation to boost the semi-supervised learning process. If the teacher can generate high-quality pseudo-labels, requirements for post-processing and post-filtering can be reduced. Our in-house dataset for structure segmentation (called \emph{Internal-Structure}) contains around 160 hours of music across a variety of styles. Our in-house dataset for downbeat and key detections (called \emph{Internal-Beat/Key}) contains 50k clips and a total size of 580 hours. This dataset covers a wide variety of genres.

For unlabeled datasets, we categorize them into three types.
\emph{Datasets of other MIR tasks (Other-MIR)}: for example, songs in downbeat datasets do not have chord annotations, so they can be part of the unlabeled dataset for chord recognition.
  %are included as an unlabeled dataset in the targeted task.
\emph{Audio-92K}: audio recordings collected internally, covering a wide range of languages, genres, and styles.
\emph{Music-151K}: music recordings collected internally from a different source than Audio-92K. It contains mostly English songs.
All the datasets used in this work are for research purposes only. %In-house datasets are de-identified with no personally identifiable information (PII). 

%\vspace{-0.6\baselineskip}

\subsection{Pseudo-labeling}
Data filtering during the teacher-student learning process is critical to the success of training a student model. In this work, we use the ``soft labels'' as the pseudo-labels. Specifically, the teacher's output probabilities are used directly as supervision signal for the student, using Cross-Entropy loss. We observe that when the teacher is more confident in a binary prediction, it tends to output a probability close to one or zero. If the teacher is less confident, it tends to output a probability around 0.5. 
Using soft labels enables the student model to learn this ``degree of uncertainty'' information from the teacher as well, so a rule-based filtering scheme is not needed.

%\vspace{-0.4\baselineskip}

\section{Experiment}\label{sec:experiment}

\subsection{Benchmark Tasks} \label{sec:benchmark}

\textbf{Downbeat Tracking} 
aims to predict the first beat of each bar, and is typically conducted together with beat tracking.
%Beat and downbeat tracking has been an important MIR task studied for years. 
In this work, we focus on downbeat tracking, because it still remains a challenge compared to beat tracking. %Our framework may benefit downbeat modeling as it requires longer context of harmonic, timbral, and structural information \cite{Hung2022ModelingBA}. 
%since the large amount of training data might help capture the harmonic context. 
Following previous works~\cite{Hung2022ModelingBA}, the model outputs frame-level probabilities of beat, downbeat, and non-beat per 50 ms. %We use one residual unit followed by a pooling size of 2 along both time and frequency dimensions to process an input harmonic-spectrogram that uses a hope size of 160 samples. The harmonic-spectrogram is extracted with a set of trainable harmonic filters \cite{Won2020DataDrivenHF}. 
A dynamic Bayesian network implemented in \texttt{madmom}~\cite{Bck2016JointBA} is used as post-processing to decode the timestamps of downbeats. 
%The DBN is implemented using madmom library. 
%Following previous work~\cite{Hung2022ModelingBA}, 
Ballroom, Beatles, Hainsworth, HJDB, RWC, Simac, Harmonix, SMC, and Internal-Beat/key are used as supervised datasets. The training process randomly samples a chunk of 6 seconds from each training song. We select 74 songs from HarmonixSet as validation set. GTZAN~\cite{Marchand2015GTZANRhythmET} is used as test set. We use F-measure implemented in \texttt{mir\_eval} ~\cite{Raffel2014MIR} for evaluation metric. 

\vspace{0.2cm}
\noindent\textbf{Chord Recognition}
is a challenging MIR task due to the lack of training data as well as the complexity of modeling the harmonic relationship in music. %We use two residual units followed by pooling sizes of 2 and 4 respectively on time and frequency dimensions to process the harmonic-spectrogram \cite{Won2020DataDrivenHF} that uses a hop size of 500 samples. 
The model outputs frame-level probabilities of 25 classes (12 pitches of major and minor chords plus one ``none'') per 125 ms. 
We use Isophonics, Billboard, HookTheory training set, RWC, and USPOP for training, and the 2000 songs of HookTheory test set for validation. 
%Since most of the previous works have very different training/testing datasets, 
We use JayChou, one of the benchmark datasets in MIREX,\footnote{https://www.music-ir.org/mirex/wiki/MIREX\_HOME} as our testing set. JayChou is not used for training by the compared models in the experiments. The evaluation metric is major/minor weighted accuracy in \texttt{mir\_eval}~\cite{Raffel2014MIR}.

\vspace{0.2cm}
\noindent\textbf{Key Detection}
aims to predict the tonal scale and pitch relation across the entire song. It has been a task in MIREX and studied for years. %We use 3 residual units followed by pooling sizes of 4 and 100 on time and frequency dimensions respectively to process the harmonic-spectrogram \cite{Won2020DataDrivenHF} that uses a hop size of 320 samples. 
The model outputs the frame-level probabilities of 25 classes (12 major and 12 minor keys plus one ``none'') per 2 seconds. We use Isophonics, Billboard, HookTheory training set, and Internal-Key/Beat for training, and the 2000 songs of HookTheory test set for validation. Giantsteps~\cite{Knees2015TwoDS} is used as test set. The evaluation metric is a refined accuracy (weighted accuracy) in \texttt{mir\_eval}~\cite{Raffel2014MIR}, with error tolerance that gives partial credits to reasonable errors.

\vspace{0.2cm}
\noindent\textbf{Structure Segmentation}
aims to partition a music recording into non-overlapping segments and predict the functional label (e.g. 'verse' and 'chorus') for each segment. Following previous work~\cite{Wang2022ToCA,wang2022musfa}, the model outputs 
frame-level label probabilities of 7 classes and a frame-level boundary probability, with a frame hop of 192 ms. We use Billboard, SALAMI, RWC, HarmonixSet, and Internal-Structure for training. 200 pieces of songs are randomly picked for validation. Isophonics is used for testing. We use the F-measure of boundary hit rate at 0.5 seconds (\emph{HR.5F} in \texttt{mir\_eval}~\cite{Raffel2014MIR}) for evaluation.

\begin{figure*}
\centering
\begin{subfigure}[b]{0.24\textwidth}
\includegraphics[width=\textwidth]{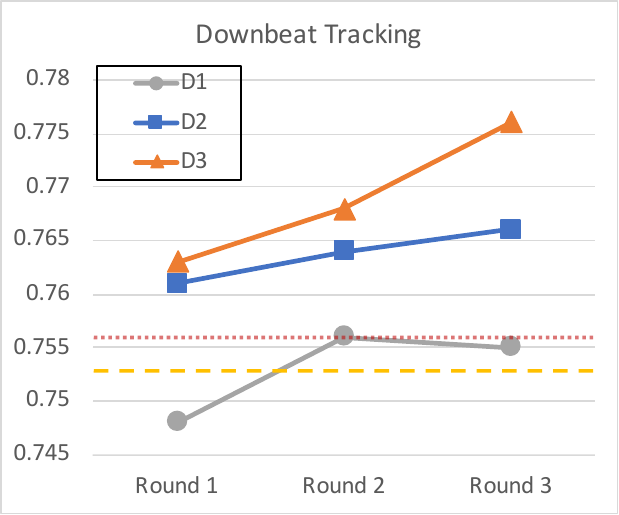}
\caption{Downbeat Tracking}
\label{fig: db_tracking}
\end{subfigure}
\begin{subfigure}[b]{0.24\textwidth}
\includegraphics[width=\textwidth]{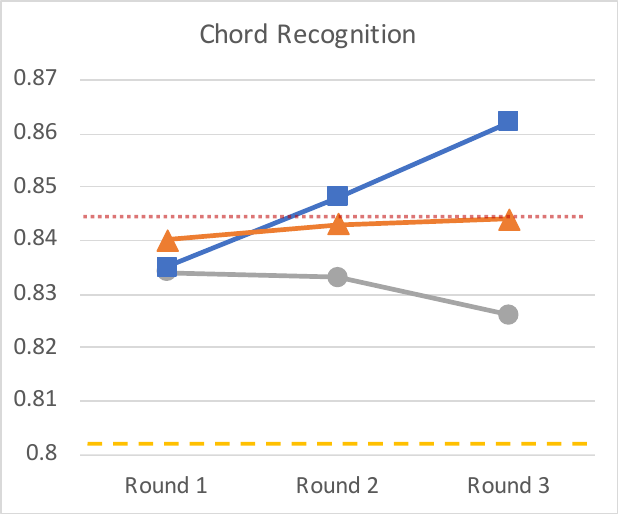}
\caption{Chord Recognition}
\label{fig: chord_recog}
\end{subfigure}
\begin{subfigure}[b]{0.24\textwidth}
\includegraphics[width=\textwidth]{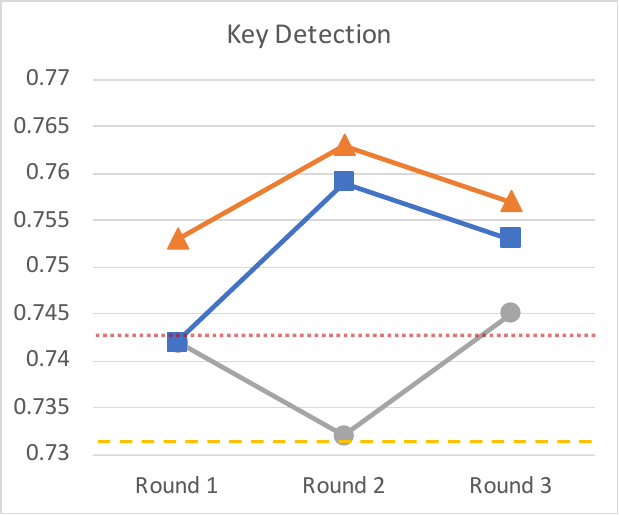}
\caption{Key Detection}
\label{fig: key_detect}
\end{subfigure}
\begin{subfigure}[b]{0.24\textwidth}
\includegraphics[width=\textwidth]{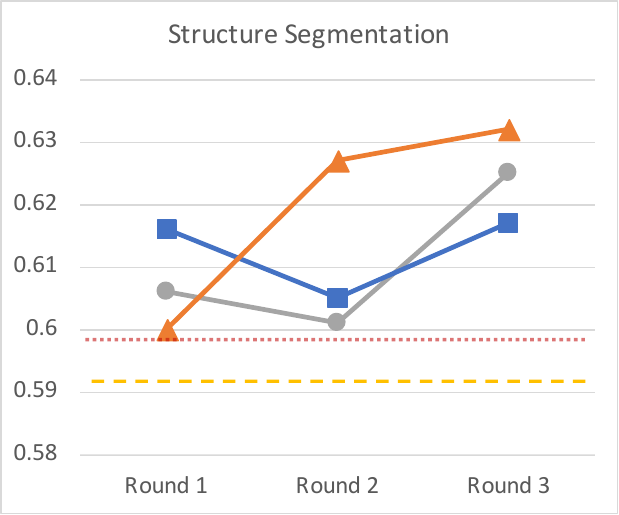}
\caption{Structure Segmentation}
\label{fig: structure_seg}
\end{subfigure}
\caption{Results of four MIR tasks with different sizes of training data and model parameters. Yellow dashed line represents the results of supervised teacher $M$. Red dotted line represents the SOTA results of non-semi-supervised training. }
\label{fig: results}
\end{figure*}

\vspace{-0.2cm}

% learning rate / batch size
\subsection{Experiment Details}
As summarized in Table~\ref{tab:parameters}, the parameters are chosen as suggested in previous works or through hyperparameters tuning based on the first round of supervised training. We use the same model architecture in the first round to train the supervised teacher model $M$. Input audio is downsampled into 16k Hz. We use Adam optimizer, 0.9 weight decay, and 20 epochs of patience. Each epoch runs 500 mini-batches, and early stopping is applied when validation scores do not decrease for 80 epochs. Model with the best validation result is used for testing. We train the models using Nvidia V100-32G GPUs. Total three rounds of training are conducted with parameters shown in Table~\ref{tab:parameters}. We consider three different data sizes for each round of training: (1) \emph{D1} is the Other-MIR dataset depending on the task; (2) \emph{D2} is D1 plus Audio-92K; (3) \emph{D3} is D2 plus Music-151K.

\vspace{-0.2cm}

\subsection{Baseline System}

We compare the proposed work with two types of baseline systems: one is the supervised teacher $M$ trained from scratch with the labeled data only; the other is based on the pre-trained model \emph{MERT-330M} \cite{Li2023MERTAM}, a state-of-the-art large-scale self-supervised model for music audio.

For the baseline systems based on MERT-330M, two training schemes are considered, namely \emph{MERT-P} and \emph{MERT-F}. In MERT-P, we freeze the MERT model to extract the deep embeddings, and then add two GRU layers for frame-level classification. We use GRU layers instead of Multilayer Perception (MLP) because they show better performance. In MERT-F, we add a dense layer (i.e., the same as the proposed model) and then fine-tune the MERT model for a fair comparison. 

\begin{table}
\centering
\small
\begin{tabular}{l c c c c}
  \hline
   & Downbeat & Key &  Chord & Structure \\
  \hline \hline
   SpecTNT-TCN \cite{Hung2022ModelingBA} & 0.756 & - & - & - \\
   %\cite{Zhao2022BeatTD} & 0.714 & - & - & - \\
   %\cite{Jiang2019Mirex} & - & 0.758 & - & - \\
   CNN \cite{Korzeniowski2017EndtoendMK} & - & 0.743 & - & - \\
   JLCX1 \cite{Jiang2018Mirex} & - & - & 0.845 & - \\
   %\cite{Park2019ABT} & - & - & 0.824 & - \\
   MuSFA \cite{wang2022musfa} & - & - & - & 0.598 \\
  \midrule 
   MERT-P & 0.736 & 0.664 & 0.723 & 0.616 \\
   MERT-F & 0.769 & 0.717 & 0.797 & --- \\
  \midrule 
  Supervised $M$ & 0.753 & 0.731 & 0.802 & 0.594 \\
  Proposed  & 0.776 & 0.764 & 0.862 & 0.632 \\
  \midrule 
 \end{tabular}
 \caption{Results for four MIR tasks. Fine-tuning MERT on structure could not fit into a V100 GPU.} 
 \label{tab:experiment}
\end{table}

%\vspace{-0.6\baselineskip}

\section{Results and Discussion} \label{sec:result}
Figure~\ref{fig: results} shows the results with different sizes of models and data on four MIR tasks. For most of the tasks, we find the performance is improved as the training round unfolds. Larger models are more likely to benefit from more data (see D2 or D3). Models trained on D3 perform the best except for chord recognition. This could be due to data mismatch: our D3 comprises mainly English songs, while JayChou contains only Chinese-pop songs. 
% Moreover, JayChou there are only 29 pieces of songs, which might not be enough to reflect the performance gains from the model. 
Nevertheless, it achieves the largest relative improvement in chord recognition as compared to its supervised teacher $M$. 
Since we do not have larger internal datasets for chord recognition, transformer could not perform well in the supervised setting. 
Among the tasks, key detection benefits the least from the proposed approach. As key is a relatively global musical event that does not vary with finer temporal resolution, there is less temporal dependency that the temporal Transformer of PerceiverTF could take advantage of. 
Due to this constraint, we also couldn't see any benefits when scaling up the model more. 
%The maximum number of model parameters for key detection is smaller (i.e., 42.8M) compared to those of other tasks.

%It is clear that the proposed models outperform their initial teacher $M$, demonstrating the effectiveness of the semi-supervised learning. The proposed models also achieve state-of-the-art performance compared to existing SOTA systems.

Table~\ref{tab:experiment} summarizes our best results compared to the baselines. 
It is clear that the proposed models all outperform their supervised teachers $M$ as well as their self-supervised counterparts, demonstrating the effectiveness of the semi-supervised framework. The proposed models also achieve state-of-the-art performance compared to existing SOTA systems, showing the proposed approach is task-independent.
MERT performs well on downbeat tracking and structure segmentation that require more temporal information, but is less successful on chord recognition and key detection that require capturing harmonic information. 
Fine-tuning MERT seems to be a better choice than probing MERT. 
The improvement is larger especially on chord recognition and key detection since it allows the pre-trained MERT to learn task-dependent information. 
% Our approach, on the other hand, learns task-specific information from the beginning, hence it outperforms MERT on all the tasks, especially on key detection and chord recognition. 
%Our proposed approach also outperforms other SOTA baselines, despite no task-specific design in our model.

%\vspace{-1\baselineskip}

\section{Conclusion}
In this paper, we investigate the effectiveness of large-scale semi-supervised MIR training by scaling up models to almost 100M parameters and training data to 240k hours.
%In this paper, we have presented the teacher-student knowledge expansion framework that leverages unlabeled data to train larger-scale models for several MIR tasks. 
%which conventionally is considered lack of training data. 
%By scaling up both model and training data, 
Addressing the challenge of scarce labeled MIR data, our experiment demonstrate that our proposed approach offers a better solution for advancing MIR performance beyond supervised or self-supervised learning approaches.
%Given the scarceness issue of labeled MIR data, we have demonstrated in the experiment that the proposed approach could offer a better solution to further the MIR performance compared to the supervised or self-supervised learning approaches. 
For future research, we aim to include more downstream tasks. Similar to other domains, we also plan to explore the impact of further scaling up the training data and model size.
%up to millions of hours and model size more to billions of parameters by incorporating large-scale training techniques.  

\newcommand{\BIBdecl}{\setlength{\itemsep}{-0.03 em}}

\bibliographystyle{IEEEtran}

\bibliography{ISMIRtemplate}

\appendix

\end{document}